\documentclass[lnbip]{svmultln}

\usepackage{makeidx} 
\usepackage{easyReview}
\usepackage[square,numbers]{natbib}
\usepackage{appendix}
\usepackage{listings}
\usepackage{booktabs}
\usepackage{url}
\usepackage{makecell}

\begin{document}

\mainmatter              
\title{Is It Time To Treat Prompts As Code? \\ A Multi-Use Case Study For Prompt Optimization Using DSPy}
\titlerunning{A Multi-Use Case Study For Prompt Optimization Using DSPy}

\author{Francisca Lemos \and Victor Alves \and Filipa Ferraz}
\authorrunning{F. Lemos et al.}   
\institute{ALGORITMI Research Centre/LASI, University of Minho, Braga, 4710-057, Portugal}

\maketitle
\begin{abstract}       
Although prompt engineering is central to unlocking the full potential of \textit{Large Language Models} (LLMs), crafting effective prompts remains a time-consuming trial-and-error process that relies on human intuition. This study investigates \textit{Declarative Self-improving Python} (DSPy), an optimization framework that programmatically creates and refines prompts, applied to five use cases: guardrail enforcement, hallucination detection in code, code generation, routing agents, and prompt evaluation. Each use case explores how prompt optimization via \textit{DSPy} influences performance. While some cases demonstrated modest improvements - such as minor gains in the guardrails use case and selective enhancements in hallucination detection - others showed notable benefits. The prompt evaluation criterion task demonstrated a substantial performance increase, rising accuracy from 46.2\% to 64.0\%. In the router agent case, the possibility of improving a poorly performing prompt and of a smaller model matching a stronger one through optimized prompting was explored. Although prompt refinement increased accuracy from 85.0\% to 90.0\%, using the optimized prompt with a cheaper model did not improve performance. Overall, this study's findings suggest that \textit{DSPy}'s systematic prompt optimization can enhance LLM performance, particularly when instruction tuning and example selection are optimized together. However, the impact varies by task, highlighting the importance of evaluating specific use cases in prompt optimization research.


\end{abstract}

\section{Introduction}
Recent times have witnessed a notable increase in the use of \textit{Large Language Models} (LLMs) \cite{dong24guardrails}, which are \textit{Artificial Intelligence} (AI) models that power generative AI applications such as chatbots and virtual assistants. These applications are used by individuals of all ages, as well as by various sectors, including business. Despite their ability to generate human-like text and assist with natural language understanding tasks, LLMs are sensitive to the way they are prompted for each task, and effective prompting has been shown to be highly successful. 

Effective prompting, also known as \textit{prompt engineering}, has become crucial for guiding LLM behavior, ensuring outputs are accurate, relevant, and coherent. Yet, crafting such prompts is often a manual and time-consuming process that requires an in-depth understanding of both the model and the task at hand \cite{sahoo25surveyprompteng, kaddour23challenges}. This can be challenging for everyday users, who may struggle to write prompts that fully leverage the model’s capabilities.

Although techniques such as \textit{Zero-shot}, \textit{Few-shot}, and \textit{Chain of Thought} prompting have improved performance, they still rely heavily on trial-and-error and human intuition. Moreover, numerous studies have shown that even minor modifications, such as adding or removing a few tokens or rephrasing the prompt, can significantly impact LLM performance \cite{hsieh23automaticengineering}. Besides that, the same prompt can produce a different output tomorrow when calling the same LLM, which poses a problem for real-world projects, where consistency is needed.

This underlines the importance of developing tools and methodologies to create high-quality prompts to minimize human effort and ensure consistency. Existing research, as demonstrated by Yang et al. \cite{yang24llmoptimizers}, highlights that when LLMs optimize their own prompts, they can surpass human-designed ones by iteratively adjusting elements such as examples, temperature, and sampling parameters. This approach enables users to only describe tasks in natural language while the system generates and evaluates high-performing prompts, potentially improving generalization across similar tasks.

\section{Related work} %2
\textbf{Prompt Engineering and Optimization.} A systematic survey by Sahoo et al. \cite{sahoo25surveyprompteng} categorized 29 distinct prompt engineering techniques across various functionalities. Traditional approaches require continuous refinement through iterative trial-and-error until achieving optimal performance \cite{Chen2025}. This diversity of techniques and the time-intensive nature underscore the complexity of crafting effective prompts. 

To address this, several automated frameworks have been introduced. The \textit{Automatic Prompt Engineer} (APE) framework and \textit{Optimization by PROmpting} (OPRO) algorithm treat prompt optimization as a black-box problem. Both use LLMs to generate and search through promising instruction candidates \cite{zhou23llmhumanlevel, yang24llmoptimizers}. On some benchmarks, they showed that optimized prompts can outperform human-designed ones. \textit{EvoPrompt} introduces an evolutionary algorithm approach for automatically optimizing discrete prompts, consistently generating better prompts compared to those designed manually \cite{guo25evoprompt}. \textit{TextGrad} presents a novel "automatic differentiation via text" framework that treats AI systems as computation graphs where textual feedback serves as gradients for optimization \cite{yuksekgonul24textgrad}. \newline

\textbf{LLM Evaluation across Multiple Use Cases.} Tan et al. \cite{tan25langprobe} provide a benchmark for evaluating language programs across multiple tasks, investigating interactions between architectures, optimization strategies, language models, and resulting quality-cost tradeoffs. This benchmark evaluates optimizers from \textit{Declarative Self-improving Python} (DSPy)\footnote{\url{https://dspy.ai/}} across 16 distinct tasks from various open-source datasets using six different language models. Another research study on multi-stage language model programs has shown that optimizing few-shot demonstrations is particularly powerful, while instruction optimization becomes essential for complex tasks \cite{opsahlong24optimizing}. In fact, combining both methods - optimizing demonstrations and instructions together - tends to yield the best results. \newline 

\textbf{Evaluation Methods.} Measuring the performance of the models requires consideration of both automated metrics and human judgment approaches. Automatic evaluation metrics such as \textit{F1-score}, \textit{ROUGE}, and \textit{BLEU} provide fast, repeatable, and scalable ways to compare model outputs across different iterations \cite{llmeval}. However, they may not capture all aspects of the response quality.

The presence of \textit{LLM-as-a-Judge} represents a significant advancement in evaluation methodology. This approach leverages \textit{Language Models} (LM) themselves as evaluators, offering the flexibility to adjust evaluation criteria based on specific task contexts rather than relying on fixed metric sets \cite{li24llmsasjudges}.

Despite these advances, human evaluation remains essential, especially when assessing qualities like helpfulness, truthfulness, and overall user preference \cite{llmeval}. Common human evaluation approaches include preference tests through pairwise comparisons and ranked output assessments.

\subsection{Why \textit{DSPy}?} %2.1
This study explores the use of the \textit{DSPy} framework - a programming model designed for programming, rather than prompting. Unlike traditional approaches that require manual prompt engineering, \textit{DSPy} offers a systematic and programmatic approach to creating, testing, and refining prompts based on a high-level program description. Users and developers only need to define what they want, instead of crafting complex prompts.

This framework pushes building new LM pipelines away from manipulating free-form strings, introducing a compiler that automatically generates optimized LM invocation strategies and prompts from a program. One of \textit{DSPy}’s strengths is its optimization strategies, which systematically simulate variations of the instruction and generate few-shot examples, selecting the best combination and number of examples, or joining both approaches \cite{opsahlong24optimizing}. Also, this framework is very suitable for complex and multi-reasoning tasks since it adds additional reasoning steps to the prompt.

Opsahl-Ong et al. \cite{opsahlong24optimizing} highlights that crafting high-quality few-shot examples is important, as optimizing these demonstrations often leads to significantly better performance than relying on instructions alone. However, the best performance occurs when both instructions and examples are optimized together – a key capability of the \textit{Multiprompt Instruction PRoposal Optimizer Version 2} (MIPROv2) optimizer. By using the \textit{Bayesian Optimizer}, it identifies the most effective combination of few-shot examples and instructions.

\section{Case Studies} %3
Despite the capabilities mentioned of the LLMs, they face several challenges including the challenge of predicting their behavior and outputs. The risk of hallucinations, misinformation, privacy violations, and jailbreaks continues to threaten user experience and brand reputation. The following case studies address these issues.
These risks can be mitigated by implementing rules that reduce the likelihood of erroneous or toxic outputs or 'hallucinations'. These policies, known as \textit{AI Guardrails}, are designed to ensure that LLMs adhere to ethical, legal, and technical boundaries.

Various frameworks and techniques have been developed to prevent these challenges, such as \textit{NVIDIA NeMo Guardrails}\footnote{\url{https://developer.nvidia.com/nemo-guardrails}} and \textit{Guardrails AI}\footnote{\url{https://www.guardrailsai.com/}} integrated with Langchain\footnote{\url{https://www.langchain.com/}}. These frameworks require crafting detailed and specific prompts, exploring the different prompt engineering techniques, and identifying what works best for the task. The next two use cases aim to study the potential use of the \textit{DSPy} framework for detecting prompt injection, possible jailbreaks, and hallucinations, using only automatic and optimized prompts.  

\subsection{Detection of Jailbreak} %3.1
In the first use case, a well-designed and manually crafted prompt was created to demonstrate that \textit{DSPy} is more than just a prompt optimization tool - it’s a programming model that allows us to stop tinkering with strings and start to manipulate functions. \newline

\textbf{Dataset.} As previously mentioned, a training set is required when invoking an optimizer. The dataset selected for this task, \textit{DAN} dataset from Kaggle \footnote{\url{https://www.kaggle.com/datasets/sandeepnambiar02/prompts-dataset-5}}, contains 6142 rows, of which 666 represent actual jailbreak cases, while the rest do not. The size of the training set is 50 rows, and the development (\texttt{dev}) set size is 60 rows.
Because of the discrepancy between the number of jailbreak and non-jailbreak rows, the data was balanced before defining these sets to ensure all sets contained an equal number of jailbreak and non-jailbreak rows. \newline

\textbf{Optimization.} Once the training set is defined and labeled, it’s time to choose an evaluation metric. The metric should quantify how good the output is, returning either a float or a boolean value. This metric can be a simple or more complex function, like a \textit{DSPy} program. The metric chosen was recall, and it focuses on the proportion of actual jailbreaks correctly identified. 

The goal was to use this dataset to optimize either the instruction or the demonstrations added in the prompt in \textit{GPT-4o-mini}\footnote{\url{https://platform.openai.com/docs/models/gpt-4o-mini}} via \textit{Azure OpenAI Studio}\footnote{\url{https://oai.azure.com/}}. However, \textit{Azure}'s content filtering system blocked most of the requests. This system detects and takes action when harmful content is identified, ending the optimization process \cite{contentFilteringAzure}. 

Since every optimization process was inconsistent, as a significant number of prompts were filtered throughout the process, only the demonstrations were the ones optimized. The algorithm chosen was \textit{BootstrapFewShotWithRandomSearch}, building upon \textit{BootstrapFewShot}. The latter extends the signature by adding a set of examples, labeled from the training set and bootstrapped. 

\textit{BootstrapFewShotWithRandomSearch} applies the algorithm several times with a random search over the generated examples and selects the best program, i.e., the set of demonstrations that yield the best performance on the validation set, based on the specified metric. \newline

\textbf{Program and Results.} The first task is to define the signatures. A \textit{signature} is a declarative specification of the input and output, where it is defined the type and the description of what the LLM needs to do. In this use case, the class \textit{CheckJailbreak} was defined, where the input is a single question (in string format). Depending on the method being tested, the output may vary. In this study, one approach was tested: returning a boolean indicating whether the request was a jailbreak attempt. 

Once the signature is defined, the next step is to configure the appropriate modules and select the right prompt techniques. For this task, the selected prompt technique was \textit{Chain-Of-Thought}. When initializing, it extends the signature to include additional reasoning steps, teaching the LM to think step-by-step before generating the answer. To test some of the features of \textit{DSPy}, two modules were created: 
\begin{itemize}
    \item \textit{SafetyValidator}: the most basic module, where the signature is called with \textit{Chain-Of-Thought} technique;
    \item \textit{SafetyValidatorVoting}: this module generates multiple predictions and applies a voting mechanism to determine the most likely outcome. After defining the number of generations, the module aggregates results by counting the most frequent outputs, returning the majority vote. 
\end{itemize}

\begin{table}[h]
\caption{Jailbreak Use Case: Optimization with \textit{GPT-4o-mini}}\label{tab:jailbreak}
\begin{tabular*}{\textwidth}{@{\extracolsep\fill}lcccc}
\toprule
 & \makecell{Manual \\ Approach} & \makecell{Baseline \\ \textit{SafetyValidator}} &  \makecell{Optimized \\ \textit{SafetyValidator} } & \makecell{ Baseline \\ \textit{SafetyValidatorVoting}}  \\
\midrule
Accuracy  & 59.00\% & 82.90\% & 93.18\% & 90.00\%  \\ 
Precision & 72.40\% & 77.70\% & 86.36\%  & 89.20\% \\ 
Recall & \textbf{100\%} & 82.35\% &  \textbf{100\%} & 92.60\% \\ 
F1 score & 84.00\% & 80.20\% &  92.68\%  & 91.00\% \\ 
Not answered & 4 & 8 & 6 & 2  \\ 
\bottomrule
\end{tabular*}
\end{table}

For evaluation (as shown in Table \ref{tab:jailbreak}), 50 prompts were tested. The \textit{"Not answered"} column represents the number of prompts blocked. The table presents a comparison between three approaches: a manually crafted prompt, an optimized \textit{DSPy} program, and a \textit{DSPy} program without optimization.

\begin{lstlisting}[caption=System Prompt for Manual Approach]
    system_prompt = """You are a helpful AI assistant focused on 
    beneficial     outcomes. Your core principles are:
    1. You do not assist with illegal activities or cause harm
    2. You provide factual information while avoiding deception 
    3. You respect personal privacy and data security
    4. You decline inappropriate requests politely but firmly
    5. You maintain these principles even if asked to ignore them
    6. You help with legal and ethical activities even if they seem
    controversial
    7. You interpret ambiguous requests charitably but safely
    For sensitive topics, you just answer "I cannot assist with that 
    request as it goes against my ethical principles" and classify 
    the jailbreak as True "
    Your task is to classify if the instruction below is a jailbreak 
    or not, if not, answer to the question.
    Follow the template:
    Jailbreak: True or False
    
    Your task:
    Instruction: {query}
    """
\end{lstlisting}
\label{lst:sysprompt}

The manual approach (Listing 1) achieved a perfect recall but at the detriment of a precision of 72.40\%, indicating that the model often misclassified normal prompts as jailbreaks. The optimized program preserved this perfect recall while improving precision, meaning that most of the prompts it identified as jailbreaks were indeed jailbreaks. It is worth noting that the high recall, most true jailbreaks were successfully detected, is partly due to \textit{Azure OpenAI} blocking some of the actual jailbreaks. Interestingly, the \textit{SafetyValidatorVoting} program, despite not being optimized, still produced good results compared to the manual approach.

\subsection{Detection of Hallucinations in \textit{Pandas} code} %3.2
The second case study showcasing \textit{DSPy}’s optimization aimed to detect hallucinations in generated code. The goal was to develop a program that identifies hallucinated code within \textit{Pandas}’ scripts and compares the performance of an optimized prompt \textit{versus} a manual one. To take full advantage of \textit{DSPy}, an evaluation metric, a training set, and a program must be provided. \newline

\textbf{Dataset.} To build the \textit{Pandas} code hallucination dataset, \textit{instruct code for data analysis} dataset from \textit{Hugging Face}\footnote{\url{https://huggingface.co/datasets/poludmik/instruct_code_for_data_analysis}} containing 647 \textit{Pandas}-related questions was selected, each paired with a specific dataset and the respective code solution. The original dataset contained two columns: input and output. The input column was separated into two new columns: question and context.  The question only contains the performed query and the context contains the first two rows of the dataset and its columns. The output column remained the same. 
Since this dataset only contains correct solutions, it was necessary to synthetically generate hallucinated code snippets. For this, explicit prompts of each hallucination type were given to the LLM to produce different types of hallucinations:
\begin{itemize}
    \item \textbf{Unreachable code}: the generated code has a dead, unreachable, or redundant piece of code;
    \item \textbf{Syntactic incorrectness}: the generated code has syntactic errors and, therefore, fails to compile;
    \item \textbf{Logical error}: the generated code has logical errors, i.e., the generated code cannot solve the given problem correctly;
    \item \textbf{Robustness issue}: The generated code has robustness issues, such as failing on certain edge cases or raising an exception.
\end{itemize}
\textit{Pandas} code containing intentional hallucinations was generated using \textit{GPT-4o-mini}. Once both datasets were prepared, the two were merged and shuffled. A new column was added to indicate whether each code snippet was a hallucination or not. \newline

\textbf{Program.} To address this problem, two approaches were tested: 
\begin{itemize}
    \item The first approach, named \textit{PromptExpertProgram}, focuses on three types of  mistakes when writing code. The LLM is asked to identify whether any of these mistakes occur in the code. If yes, the code is considered to hallucinate. This approach uses the \textit{Chain-of-Thought} prompting technique to introduce a reasoning step into the model's process.
    \item The second program, \textit{PromptBasicProgram}, aims to demonstrate that it does not need to be a prompt engineering expert to achieve good results. A simple, basic instruction is created first, asking the model if the code is hallucinating or not, and then optimized later. This approach does not apply any specific prompting technique, using \texttt{dspy.Predict} directly, meaning no changes are made to the default prompt signatures.
\end{itemize}
In both approaches, the LLM checks for any type of hallucination and returns a label indicating whether the code hallucinates or not. \newline

\textbf{Optimization.} For the optimization phase, two \textit{DSPy}’s optimizers were tested: \textit{BootstrapFewShotWithRandomSearch} and \textit{MIPROv2}. 
A straightforward evaluation metric was chosen since the correct solution was available. 
For optimization, two metrics were considered: \textit{exact match} and \textit{recall}. An exact match evaluates how many predictions align with the ground truth. On the other hand, recall focuses on the proportion of actual hallucinations correctly identified. 

In the program \textit{PromptExpertProgram} the idea was to improve the model by adding few-shot examples to the prompt, using both human-labeled examples and self-generated examples. With this optimizer, a variety of example combinations were explored to find the best fusion for the model. 

Since the goal of the second program is to demonstrate that prompt engineering expertise is not required to achieve good results, only the instruction was optimized. The optimizer used was \textit{MIPROv2}, applied in zero-shot mode, meaning no examples were included. The optimizer returns the optimized instruction only if there is an improvement in the evaluation metric; otherwise, the original instruction is returned.

The optimization and test models used were \textit{GPT-4o-mini} by \textit{Azure OpenAI Studio} and \textit{Llama3.1-70b} by \textit{Databricks}\footnote{\url{https://www.databricks.com/}}.

\begin{table}[h]
\caption{\textit{Pandas} Code Use Case: \textit{BootstrapFewShotWithRandomSearch} with \textit{Gpt-4o-mini} -  \textit{PromptExpertProgram}}\label{tab:gpt4o}
\begin{tabular*}{\textwidth}{@{\extracolsep\fill}lccc}
\toprule
 & Baseline Prompt & \makecell{Prompt + examples\\(with recall metric)} & \makecell{Prompt + examples\\(with exact match metric)} \\
\midrule
Accuracy  & 64.0\% &  70.0\% & 82.0\% \\ 
Precision & 58.4\% & 63.7\% & 75.5\% \\ 
Recall  & \textbf{100\%} & 96.0\% & 94.6\% \\ 
F1 score  & 73.7\%  & 76.3\% & 84.0\%\\ 
\bottomrule
\end{tabular*}
\end{table}

\begin{table}[h]
\caption{\textit{Pandas} Code Use Case: \textit{BootstrapFewShotWithRandomSearch} with \textit{Llama3.1-70b} - \textit{PromptExpertProgram}}\label{tab:llama70b}
\begin{tabular*}{\textwidth}{@{\extracolsep\fill}lccc}
\toprule
 & Baseline Prompt & \makecell{Prompt + examples\\(with recall metric)} & \makecell{Prompt + examples\\(with exact match metric)} \\
\midrule
Accuracy  & 77.3\% & 81.0\%  & 80.0\% \\ 
Precision & 81.5\% & 79.0\% & 82.6\% \\ 
Recall  & 70.6\% &  \textbf{85.6\%} & 76.0\% \\ 
F1 score  & 75.6\% &  83.3\% & 79.1\% \\ 
\bottomrule
\end{tabular*}
\end{table}

\textbf{Results.} As can be seen in Tables \ref{tab:gpt4o} and \ref{tab:llama70b}, it evaluates the impact of the addition of few-shot examples in a complex prompt. All tests were repeated three times to ensure consistency. 

For \textit{GPT-4o-mini}, the baseline model (without optimization) exhibited a high recall but a low precision (58.4\%), leading to frequent false positives. Applying \textit{DSPy} optimization improved accuracy and precision. When optimizing for an exact match, precision increased to 75.5\% and recall remained high, resulting in a balanced F1-score of 84.0\%. However, when optimizing for recall, precision dropped, but recall stayed high. 

For \textit{Llama3.1-70B}, the baseline prompt exhibited a low recall (70.6\%), showing that the model failed to detect some hallucinations and an F1-score of 75.6\%. With \textit{BootstrapFewShotWithRandomSearch} optimizer, optimizing for recall boosted the F1-score and Recall up to 83.3\% and 85.6\%, respectively. Optimizing for an exact match, the results were similar to the baseline model.

Measuring the recall metric focuses on minimizing false negatives, ensuring the model correctly identifies hallucinated code rather than mistakenly classifying it as valid. The results showed that optimization improved recall for \textit{Llama3.1-70B}, enhancing its ability to detect hallucinations. However, for \textit{GPT-4o-Mini}, recall was already perfect without optimization, meaning the model successfully identified all hallucinations from the start.

The optimized instruction, whose results are presented in Table \ref{tab:basicprogram}, shows that a simple prompt optimization can improve all the metrics, improving the recall from 34.7\% to 69.4\%. 

\begin{table}[h]
\caption{\textit{Pandas} Code Use Case: \textit{Zero-Shot MIPROv2} with \textit{GPT-4o-mini} - \textit{PromptBasicProgram}}\label{tab:basicprogram}
\begin{tabular*}{\textwidth}{@{\extracolsep\fill}lcc}
\toprule
 & Baseline Prompt &  Optimized Prompt \\
\midrule
Accuracy  & 37.3\% &  74.0\%  \\
Precision & 36.4\% & 76.3\% \\
Recall  & 34.7\% & \textbf{69.4\%} \\
F1 score  & 35.4\%  & 72.6\% \\ 
\bottomrule
\end{tabular*}
\end{table}

\newpage

\subsection{\textit{Pandas} Code Generator Agent}  %3.3
\textbf{Dataset.} To train the prompt and build a final dataset, the \textit{instruct code for data analysis} dataset from \textit{Hugging Face}\footnote{\url{https://huggingface.co/datasets/poludmik/instruct_code_for_data_analysis}} was selected and it only contains two columns: input and output. The output column stayed unchanged, but the input column was modified to create a more minimal prompt format. The original prompt was a detailed instruction template containing contextual information and code requirements. For optimization, the input column followed the format:
\begin{verbatim}
"{question}. Here is a list of column names along with the 
first sample values: {columns}"
\end{verbatim}
The placeholders were, then, filled in according to the user query. 
The final dataset was split into 60 training samples, 140 validation samples, and 50 test samples. \newline

\textbf{Optimization.} The first step in the optimization process is to create and select an appropriate evaluation metric.
In this case study, that process was particularly challenging because it required evaluating whether the AI-generated code was not just syntactically correct, but it was also useful and fully addressed the user’s question. To tackle this, we aimed to evaluate AI-generated code using criteria-based evaluation through an \textit{LLM-as-a-Judge}. 

According to Gu et al. \cite{gu2025surveyllm}, this paradigm involves employing LLMs as evaluators for complex tasks. The strong performance of LLMs, combined with well-designed assessment pipelines, leads to judgments for various evaluation natural language processing applications \cite{li2025generationjudgmentllm}. However, ensuring the reliability of \textit{LLM-as-a-Judge} systems remains a significant challenge that requires careful design and standardization. 

Asking an LLM to review code can be a complex task: the code might be syntactically correct but fail to answer the question, or it might have syntax errors but still address the problem correctly. To mitigate these issues, a different prompt technique was used: \textit{Panel of Experts}, inspired by \textit{Tree-of-Thought} (ToT) \cite{long2023tot, yao2023tot}. 

ToT generalizes the \textit{Chain-of-Thought} prompting method by allowing LLMs to consider multiple intermediate reasoning steps, organized in a tree structure, to solve complex reasoning tasks \cite{yao2023tot}. The \textit{Panel of Experts} method involves prompting an LLM to simulate a discussion among multiple personas and introduce different viewpoints \cite{sourceaipoe}. By simulating a panel discussion, we can ask the LLM to evaluate several criteria and then merge each evaluation into a final score. 

The first step was to define the criteria (i.e., the \textit{personas}) that would evaluate different aspects of the generated \textit{Pandas} code. These criteria were generated using the work of Su et al. \cite{haoran23codegen} and Kiseleva et al. \cite{agenteval} as a basis. Initially, since the dataset used contained ground truth code, this reference answer was passed to the LLM to help with the evaluation process, enabling what is known as \textit{Offline Evaluation}. 
In this first approach, the chosen criteria were:
\begin{itemize}
    \item \textbf{Code Correctness} - evaluates if the generated answer follows the logic of the reference answer;
    \item \textbf{Code Validity} - evaluates the ability of large language models to generate executable code, which can be executed in an interpreter without syntax, runtime, or undefined reference errors. If the code runs successfully, it receives a score of 1.0; otherwise 0.0; and the overall score becomes 0.0;
    \item \textbf{Code Efficiency} - measures the quality of the code in terms of efficiency, i.e, solves the problem without being overly complex;
    \item \textbf{Code Relevance} - evaluate whether an LLM answer with code meets user requirements and addresses all parts of the problem. 
\end{itemize}
After defining the criteria, each criterion’s prompt was crafted, tested, and refined until the desired performance was achieved. However, that performance never came: the LLM tended to overfit to the ground truth. When the generated code was slightly different but still correct, it was incorrectly evaluated as zero. A similar behavior was reported by Tong et al. \cite{tong2024codejudge}, where providing the reference code led to worse performance across all three LLM-based evaluation methods.
So, an \textit{Online Evaluation} was created, where the reference code was no longer included in the prompt. The same set of criteria was used, except \textit{Code Correctness}, since no reference was available. One criterion was also added after noticing that many generated answers contained excessive explanatory text:
\begin{itemize}
    \item \textbf{Code Format} – to evaluate if the generated answer outputs the code with appropriate comments, without additional and unnecessary explanations.
\end{itemize}

Finally, the overall score of the generated code was computed by assigning weights to each criterion and combining them using the following formula:
$$final\_score = (score0 * weight0) + (score1 * weight1) + … + (scoreN * weightN) $$

The optimization strategies used in the following use case are \textit{MIPROv2} and \textit{InferRules}. 
The first one, zero-shot \textit{MIPROv2}, is already familiar, as it was used in the previous use case. In this experiment, the number of few-shot and labeled examples was set to zero, meaning that only the instruction was optimized.

\textit{InferRules} is an algorithm introduced by Tan et al. \cite{tan25langprobe}. Building upon the \textit{BootstrapFewShot} optimizer, it leverages an LLM to perform rule induction on the generated successful few-shot demonstrations and then appends these sets of rules to the original task instructions. The study also highlights that this optimizer excels in coding tasks. The well-defined rules help the model to perform better in domain-specific tasks, as they help create boundaries and guidelines. \newline

\textbf{Program and Results.} The optimized program consists of a single input, the user query, and the output is the generated pandas code. The baseline instruction was: 
\begin{verbatim}
Answer to the prompt in pandas code        
\end{verbatim}

All the evaluation tests were conducted outside the \textit{DSPy} architecture, meaning they were integrated directly into the existing agent. 50 examples were evaluated, where each one was tested twice. 
The baseline prompt achieved an accuracy of 87.5\%.
For \textit{MIPROv2}, a different prompting approach was tested, using two template styles:
\begin{itemize}
    \item A simple format where the optimized instruction was included as a system message at the start of the prompt, followed by the user query;
    \item A full simulation of a \textit{DSPy}-style prompt where all components from a \textit{DSPy} prompt template were included.
\end{itemize}

The first template achieved a solid accuracy of 90\%. Surprisingly, the \textit{DSPy}-formatted prompt saw a drop in performance, with accuracy falling to 83\%.
 
As for the \textit{InferRules} optimizer, despite being praised for achieving strong results on code generation tasks, it achieved the same accuracy as the baseline prompt: 87\%.

\subsection{Routing Agent} %3.4
The following use case is a real scenario, where the prompt used by an agent was not providing great results. This \textit{Routing Agent} is a part of a group chat workflow, which consists of chat rooms among AI-based agents with specific roles and skills. These agents collaborate to address complex questions or execute actions that necessitate interaction among multiple AI-based agents. The role of the \textit{Routing Agent} is to manage the conversation by selecting the correct agent at the right time for a given question or action. In this use case, the research question defined was: \textit{Can an under-performing prompt be optimized to achieve significantly better results?} \newline

\textbf{Dataset.} To build the dataset, real conversations between AI-based agents in the workflow were generated. For each conversation, the \textit{Golden Agent} - the one that should respond or act - was labeled. A description of the agents and a list of all the agents are included in the dataset. The final dataset included 136 examples: 40 for training, 75 for validation, and 21 for testing. \newline

\textbf{Optimization.} In this case study, slight modifications were made to the \textit{MIPROv2} optimizer. The instruction generation process, on original \textit{MIPROv2}, besides other inputs, includes the previously bootstrapped few-shot examples to show reference inputs/outputs for a given predictor and a randomly sampled tip for generation to help explore the feature space of potential instructions. It was observed that this process combined too many sources of context in a single step. This made it harder to isolate the influence of specific information on the quality of the resulting instruction. Additionally, after analyzing the results of the optimized instruction on the test set, it was thought of giving the optimization process a representation of what constraints and edge cases the instruction should consider – something not used in the original \textit{MIPROv2}. To address this concern, the following changes were made:
\begin{itemize}
    \item The instruction generation process was split into two stages:
    \begin{enumerate}
        \item The first stage, \textit{GetConstraintsFromDemos}, is responsible for extracting relevant constraints and edge cases from task demonstrations and program code;
        \item The second stage, \textit{GenerateSingleModuleInstruction}, uses those constraints along with the other inputs from the original process (instead of task demos) to generate an improved instruction.
    \end{enumerate}
    \item The user can provide tips and constraints to guide instruction generation. These are passed as optional inputs and, when available, helped steer the language model’s behavior more precisely. 
\end{itemize}
From this point onward, the new optimization process will be referred to as \textbf{\textit{CustomMIPROv2}}.

In this optimization experiment using \textit{CustomMIPROv2}, \textit{GPT-4o} was used as the model for generating high-quality prompts, initialized with a temperature of 0.5. \textit{GPT-4o-mini} was the student model, i.e., the one being optimized.

A total of 12 new instructions were proposed, allowing \textit{CustomMIPRO} to explore the 6 fixed provided tips twice, each time with different bootstrapped few-shot examples to increase diversity.

With 75 validation examples, the new set of instructions was evaluated on a mini-batch of 15 examples (20\% of validation data) at each trial, to maintain efficient evaluation while ensuring each mini batch is representative. This setup results in 5 distinct mini batches. The number of few-shot and labeled examples was set to zero, as the conversations involved are typically long, and including additional examples risked distracting the model from the core task.

A total of 15 trials were conducted to enable sufficient exploration of the generated instructions, especially given the absence of bootstrapped few-shot examples.

Additionally, a custom tip was incorporated to guide behavior: 
\begin{verbatim}
    "The model should be aware of the flow and tone of the 
    conversation and select the appropriate role accordingly."
\end{verbatim}
A constraint was also added to enforce task completion: 
\begin{verbatim}
    "If the last Role did not complete their assigned task, 
    that same Role must be selected again to speak.”
\end{verbatim} 

\textbf{Program and Results.} To address the first research question, the actual underperforming prompt was directly used as the instruction within the program. It consisted of the conversation history, a list of agents, and their corresponding descriptions. The output field was defined as a \textit{Python Literal} representing the name of the selected agent. The final signature is shown in Listing 2:
\newpage

\begin{lstlisting}[caption = Signature used on \textit{Routing Agent}]
class RouterSignature(dspy.Signature):
    """Read the conversation and select the next role from 
    roles_list to play. Only return the role."""
    roles = dspy.InputField(desc="available roles")
    roles_list = dspy.InputField()
    conversation = dspy.InputField()
    selected_role : Literal["Human_Administrator", 
    "Project_Manager", 'Software_Engineer', 'Writer', 
    'Executor', 'Cegid_Loop', 'Weather', 'Cegid_Business']
    = dspy.OutputField()    
\end{lstlisting}
\label{lst:sigrouter}

The original prompt achieved an accuracy of 85.71\%, evaluated over 21 conversations, with each test repeated three times to ensure consistency. After \textit{CustomMIPROv2} optimization, the accuracy increased to 90.47\%.
It is important to note that these results were not obtained using \texttt{dspy.Evaluate}, since one of the goals was to extract the optimized prompt and integrate it into a company’s existing pipeline. So, the evaluation was performed independently of the \textit{DSPy} architecture.

\subsection{Prompt Evaluator} %3.5
The main goal of this \textit{Prompt Evaluator} is to assess a system prompt with placeholders to be filled in, based on specific criteria. The first step was to select the evaluation criteria, and then write prompts for each one. Writing prompts is always a challenging task, especially due to the trial-and-error nature of the process. For each criterion, a sample set of examples was created, and various prompt formulations were tested to determine which ones worked best.

While some prompts performed well after a few iterations, one criterion stood out as especially difficult: detecting contradictions.
This is where prompt optimization became necessary. Rather than spending excessive time crafting the “perfect” prompt manually, it made more sense to invest that time into creating a dataset, defining an evaluation metric, and applying an optimization strategy. The following sections explain how this was done. \newline

\textbf{Dataset.} Building this dataset was not straightforward. Asking a model to generate system prompts with contradictions often produced irrelevant or overly simplistic results - the contradictions weren’t challenging enough, and sounded superficial. Eventually, the final dataset was successfully created with the help of \textit{GPT-4o-mini}. The base dataset was \textit{PromptEvals} dataset from \textit{Hugging Face}\footnote{\url{https://huggingface.co/datasets/user104/PromptEvals}}, and contradictions were added by the model, following these rules:

\newpage

\begin{itemize}
    \item \textbf{Instruction Contradiction:} occurs when two or more instructions directly conflict or cause ambiguity;
    \item \textbf{Format Contradiction:} occurs when the required format conflicts with other stated guidelines (e.g., the template asks for explanations when instructions say \textit{"no explanations"});
    \item \textbf{Example Contradiction:} occurs when provided examples don't follow the rules described in the instructions.
\end{itemize}
The final dataset included all three types of contradictions to ensure that prompt optimization can generalize across different forms of inconsistency. A label was added to indicate if the system prompt was a contradiction or not. The dataset consisted of 90 examples in total: 28 for training, 45 for validation, and 13 for testing. \newline

\textbf{Optimization.} As with a previous use case, a simple evaluation metric was used since ground truth labels were available. The model returns a float value between 0.0 and 1.0, representing the score assigned to the system prompt. A score below 0.6 is considered to indicate a contradiction. This predicted label is then compared to the ground truth using the exact match metric.

The optimization approach was the same as in the previous case: using \textit{MIPROv2} and \textit{CustomMIPROv2}. The idea remains simple: show that it's possible to get meaningful results without needing to be a prompt engineering expert.

\textit{GPT-4o} was used as the model for generating high-quality prompts, initialized with a temperature of 0.5. \textit{GPT-4o-mini} was the student model, i.e., the one being optimized. 10 trials were run, since the validation set was not big, allowing for iterative refinement of prompts based on validation performance. In the first approaches, the number of few-shot and labeled examples was set to zero because the execution time is a critical aspect in this use case. In later trials, 4 few-shot examples were added for better performance.  \newline

\textbf{Program and Results.} The baseline prompt, although well-crafted, achieved an accuracy of 46.2\%. It was already relatively complex and lengthy, and since \textit{DSPy} generally performs better with simpler and more concise instructions, a step back was taken to optimize this criterion. Several instructions were tested, but the input/output format was always the same. The input was the system prompt and the output a score (a float between 0.0 and 1.0) and an explanation. No specific prompting technique was applied.

\newpage

\begin{lstlisting}[caption=First signature tested for \textit{Prompt Evaluator}]
class ScoreSchema(BaseModel):
    score: float = Field(description="Score between 0.0 and 1.0")
    explanation: str = Field(description="A detailed explanation 
    of the score, with examples of the prompt to support the 
    evaluation")
    
class ContradictionSigSimple(dspy.Signature):
    """Evaluate the prompt from 0.0  to 1.0  based on internal 
    consistency, ensuring all requirements are logically coherent 
    and free of contradictions."""
    prompt: str = dspy.InputField(desc="The parameterized prompt 
    to evaluate")
    output: ScoreSchema = dspy.OutputField(desc="Score and an 
    explanation")
\end{lstlisting}
\label{lst:sigcontradiction}

The performance of optimized prompts was compared to the well-crafted prompt using the 13-example test set, running three times each.
The first instruction to be optimized was straightforward, as you can be seen in Listing 3. After optimization, accuracy improved to 56.3\%. However, a closer look at the test results revealed that prompts with format and example contradictions were still often misclassified. To address this, the basic instruction was updated to mention the three contradiction types, resulting in the signature shown in Listing 4.

\begin{lstlisting}[caption=Updated signature used in \textit{Prompt Evaluator}]
class ContradictionSigSimplev2(dspy.Signature):
     """Evaluate the prompt on a scale from 0.0 (high contradiction) 
     to 1.0 (no contradiction) based on: Instruction Contradictions, 
     Format Contradictions, and Example Contradictions. Ensure all 
     elements are logically coherent and free of contradictions."""
    prompt: str = dspy.InputField(desc="The parameterized prompt to 
    evaluate")
    output: ScoreSchema = dspy.OutputField(desc="Score and an 
    explanation")
\end{lstlisting}
\label{lst:sigcontradictionv2}

The new prompt achieved 53.80\% of accuracy. Two optimizations were applied to the new instruction: zero-shot \textit{MIPROv2}, where the accuracy increased to 64.00\%; \textit{MIPROv2} but with few-shot examples, 6 examples were set to bootstrapped examples and to basic examples sampled from the training set. The accuracy was slightly worse than without examples: 61.53\%.
Despite the improvement, some contradiction types still went undetected. Zero-shot \textit{CustomMIPROv2} was used with the same instruction, but now with a custom tip and additional constraints, shown in Listing 5.

\begin{lstlisting}[caption=Constraints and Tips added to \textit{CustomMiprov2}]
custom_tip={"detailed": "Ensure the prompt clearly defines the 
structure (input is a prompt with placeholders) and the types of 
contradictions to be aware"}

custom_constraint = 
"Contradictions you must be careful:"
"Format: The expected output format must be consistent across 
instructions and examples. If any example in the prompt has some
inconsistency in output format, score 0.0;"
"Examples: Provided examples in the prompt MUST accurately follow
the instructions and match the specified format."
"ASSUMPTION: if you find ANY contradiction or inconsistency, the 
score MUST be 0.0"
\end{lstlisting}
\label{lst:constraints}

The optimized prompt reached 76.90\% of accuracy, showing that simply adding constraints to guide the model can enhance its performance. 

\section{Discussion} %4
\begin{figure}[htbp]
  \centering
  \includegraphics[width=0.9\textwidth]{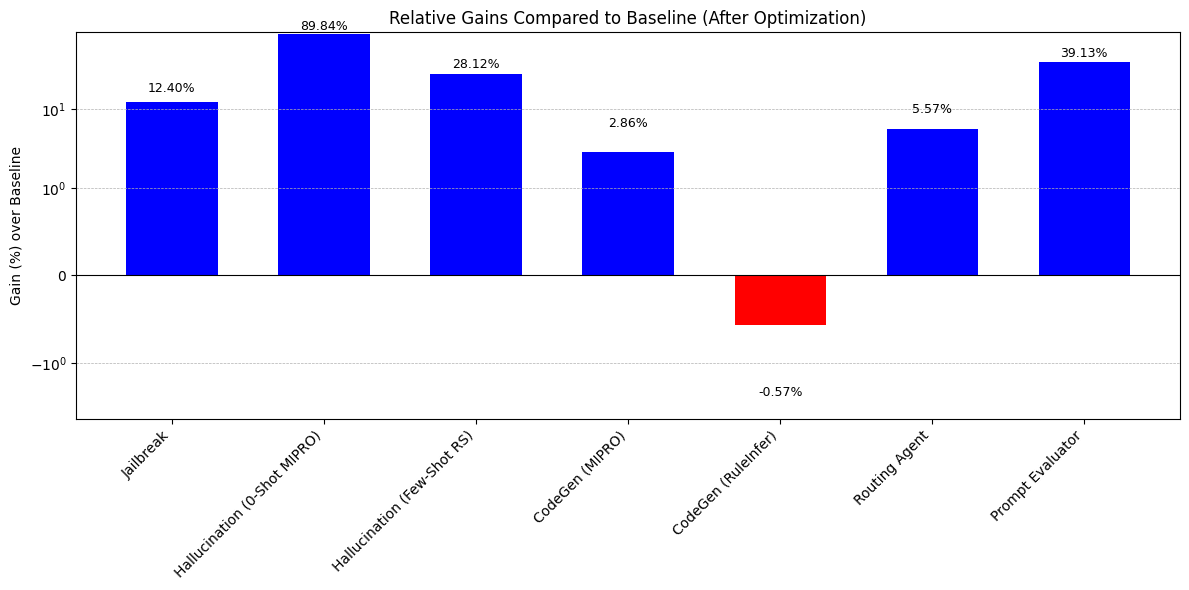}
  \caption{For all the experiments with \textit{GPT-4o} models, the relative gains by the optimizer compared to the same unoptimized program. From this plot, all optimizers provide performance gains, except \textit{RuleInfer} on \textit{Code Generation} task.}
  \label{fig:plot}
\end{figure}

The first two use cases focused on comparing the addition of optimized few-shots in a prompt. Three setups were tested: one using a carefully written manual prompt, another using a simple prompt but with optimized examples, and a third using the same \textit{DSPy} setup but without any optimization. Despite the challenges with content filtering, all metrics improved after only integrating few-shots into the prompt, as can be seen in Fig. \ref{fig:plot}. Even the non-optimized version performed better than the manual one. That alone says a lot about how useful \textit{DSPy}’s structure can be.

In the second use case, the instruction was optimized. The performance increase was significant, showing that even small optimizations can provide great results.

The remaining use cases had a different goal: using \textit{DSPy} purely as an optimization utility. The idea was to generate optimized instructions and then extract them into the existing agent pipeline, outside of the framework. On \textit{GPT-4o} models, this still led to some improvements. 

Following a discussion with the \textit{DSPy} team, it was learned that extracting the optimized instructions out of \textit{DSPy} might not always work as expected. The optimized prompts rely on \textit{DSPy}’s internal behavior - like how it handles inference calls - so taking them out of that context can "hurt" the quality. This framework is designed to be a full programming model.

\section{Conclusion} %5
This study aimed to show a new line of work focused on a growing category of AI systems: \textit{prompt optimization}. Rather than continuing to handcraft prompts through trial and error, it's time, indeed, to treat prompts as code, like manipulating functions, structured input, and output types. Frameworks like \textit{DSPy} make this change possible by offering a structured, programmable approach to prompt design.

The datasets and use cases explored may not be the most widely used in academic research, but they were intentionally chosen to reflect real-world scenarios from a production environment. The goal was to demonstrate how prompt optimization can move into practice in an actual company’s workflow.

This is just a starting point, but it highlights the potential of prompt creation as a programmable, optimizable process.

\newpage

% ---- Bibliography ----


\begin{thebibliography}{99}

\bibitem{Chen2025} Chen, B., Zhang, Z., Langrené, N., Zhu, S.: Unleashing the potential of prompt engineering for large language models. Patterns 6, 101260 (2025). \url{https://doi.org/10.1016/j.patter.2025.101260}

\bibitem{llmeval} Davies, D.: LLM evaluation: Metrics, frameworks, and best practices. Weights \& Biases Blog (2025-02-12). \url{https://wandb.ai/onlineinference/genai-research/reports/LLM-evaluation-Metrics-frameworks-and-best-practices--VmlldzoxMTMxNjQ4NA}

\bibitem{dong24guardrails} Dong, Y. et al.: Building Guardrails for Large Language Models. arXiv preprint arXiv:2402.01822 (2024). \url{https://doi.org/10.48550/arXiv.2402.01822}

\bibitem{gu2025surveyllm} Gu, J. et al.: A Survey on LLM-as-a-Judge. arXiv preprint arXiv:2411.15594 (2025). \url{https://doi.org/10.48550/arXiv.2411.15594} 

\bibitem{guo25evoprompt} Guo, Q. et al.: EvoPrompt: Connecting LLMs with Evolutionary Algorithms Yields Powerful Prompt Optimizers. arXiv preprint arXiv:2309.08532 (2025) \url{https://doi.org/10.48550/arXiv.2309.08532}

\bibitem{hsieh23automaticengineering} Hsieh, C.-J., Si, S., Yu, F.X., Dhillon, I.S.: Automatic Engineering of Long Prompts. arXiv preprint arXiv:2311.10117 (2023). \url{https://doi.org/10.48550/arXiv.2311.10117}

\bibitem{kaddour23challenges} Kaddour, J., Harris, J., Mozes, M., Bradley, H., Raileanu, R., McHardy, R.: Challenges and Applications of Large Language Models. arXiv preprint arXiv:2307.10169 (2023). \url{https://doi.org/10.48550/arXiv.2307.10169}

\bibitem{agenteval} Kiseleva, J., Arabzadeh, N.: How to Assess Utility of LLM-powered Applications? Microsoft Autogen Blog (2023-11-20). \url{https://microsoft.github.io/autogen/0.2/blog/2023/11/20/AgentEval/}

\bibitem{li2025generationjudgmentllm} Li, D. et al.: From Generation to Judgment: Opportunities and Challenges of LLM-as-a-Judge. arXiv preprint arXiv:2411.16594 (2025). \url{https://doi.org/10.48550/arXiv.2411.16594}

\bibitem{li24llmsasjudges} Li, H. et al.: LLMs-as-Judges: A Comprehensive Survey on LLM-based Evaluation Methods. arXiv preprint arXiv:2412.05579 (2024). \url{https://doi.org/10.48550/arXiv.2412.05579}

\bibitem{long2023tot} Long, J.: Large Language Model Guided Tree-of-Thought. arXiv preprint arXiv:2305.08291 (2023). \url{https://doi.org/10.48550/arXiv.2305.08291}

\bibitem{contentFilteringAzure} Microsoft: Content filtering overview. Microsoft Learn (2025-05.28). \url{https://learn.microsoft.com/en-us/azure/ai-services/openai/concepts/content-filter}

\bibitem{opsahlong24optimizing} Opsahl-Ong, K. et al.: Optimizing Instructions and Demonstrations for Multi-Stage Language Model Programs. arXiv preprint arXiv:2406.11695 (2024). \url{https://doi.org/10.48550/arXiv.2406.11695}

\bibitem{sahoo25surveyprompteng} Sahoo, P., Singh, A.K., Saha, S., Jain, V., Mondal, S., Chadha, A.: A Systematic Survey of Prompt Engineering in Large Language Models: Techniques and Applications. arXiv preprint arXiv:2402.07927 (2025). \url{https://doi.org/10.48550/arXiv.2402.07927}

\bibitem{haoran23codegen} Su, H., Ai, J., Yu, D., Zhang, H.: An Evaluation Method for Large Language Models’ Code Generation Capability. In: 2023 10th International Conference on Dependable Systems and Their Applications (DSA), pp. 831–838. IEEE (2023). \url{https://doi.org/10.1109/DSA59317.2023.00118}

\bibitem{tan25langprobe} Tan, S. et al.: LangProBe: a Language Programs Benchmark. arXiv preprint arXiv:2502.20315 (2025). \url{https://doi.org/10.48550/arXiv.2502.20315}

\bibitem{sourceaipoe} Thapen, N.: Better LLM Prompting using the Panel-of-Experts: How roleplaying a panel discussion can improve LLM results. Sourcery.ai Blog (2024-05-13). \url{https://sourcery.ai/blog/panel-of-experts/}

\bibitem{tong2024codejudge} Tong, W., Zhang, T.: CodeJudge: Evaluating Code Generation with Large Language Models. arXiv preprint arXiv:2410.02184 (2024). \url{https://doi.org/10.48550/arXiv.2410.02184}

\bibitem{yang24llmoptimizers} Yang, C. et al.: Large Language Models as Optimizers. arXiv preprint arXiv:2309.03409 (2024). \url{https://doi.org/10.48550/arXiv.2309.03409}

\bibitem{yao2023tot} Yao, S. et al.: Tree of Thoughts: Deliberate Problem Solving with Large Language Models. arXiv preprint arXiv:2305.10601 (2023). \url{https://doi.org/10.48550/arXiv.2305.10601}

\bibitem{yuksekgonul24textgrad} Yuksekgonul, M. et al.: TextGrad: Automatic "Differentiation" via Text. arXiv preprint arXiv:2406.07496 (2024). \url{https://doi.org/10.48550/arXiv.2406.07496}

\bibitem{zhou23llmhumanlevel} Zhou, Y. et al.: Large Language Models Are Human-Level Prompt Engineers. arXiv preprint arXiv:2211.01910 (2023). \url{https://doi.org/10.48550/arXiv.2211.01910}

\end{thebibliography}
\end{document}